\documentclass{ws-procs975x65}

\begin{document}



\title{kHz QPO pairs expose the neutron star of Circinus X-1
}

\author{S. BOUTLOUKOS}

\address{Center for Theoretical Astrophysics, Department of Physics, University of Illinois at Urbana-Champaign, 1110 W Green, 61801, Urbana, IL, USA
\email{stratos@uiuc.edu}
}

\author{M. VAN DER KLIS, D. ALTAMIRANO, M. KLEIN-WOLT, R. WIJNANDS
}

\address{Astronomical Institute, University of Amsterdam,
Kruislaan 403, 1098 SJ, Amsterdam, the Netherlands\\
}




\begin{abstract}
We discovered kHz QPOs in 80 archived RXTE observations from the peculiar low-mass X-ray binary (LMXB) Circinus X-1. In 11 cases these appear in pairs in the frequency range of $\sim$230 Hz to $\sim$500 Hz for the upper kHz QPO and $\sim$56 Hz to $\sim$225 Hz for the lower kHz QPO.
Their correlation with each other, which is similar to that of frequencies of kHz QPO pairs in other LMXBs containing a neutron star, and their variation by a factor two
confirm that the central object is a neutron star.
These are the lowest frequencies of kHz QPO pairs discovered so far and extend the above correlation over a frequency range of factor four.
In this new frequency range the frequency difference of the two kHz QPOs increases monotonically by more than $\sim$170 Hz with increasing kHz QPO frequency, challenging theoretical models.
\end{abstract}

\bodymatter

\section{Introduction}\label{intro}
Quasi periodic-oscillations (QPOs) are thought to reflect the motion of matter 
in the curved spacetime around neutron stars and black holes. The high frequency
QPOs have proven to be a good indicator of the nature of the compact object.
If the LMXB includes a neutron star, pairs of kilohertz (kHz)
QPOs often appear simultaneously with frequencies of several hundred Hz.
Their frequencies vary as the system evolves but their difference
appears to reflect
the spin frequency of
the star or half of it, in case where this is known from X-ray pulsations.
Their coherences and amplitudes range up to about 200 and 20\% respectively.
On the other hand, black holes only rarely show high frequency QPOs \footnote{The term kHz QPOs does not apply to them since they never reach 1~kHz.}
and then mostly one at a time; the frequency ratio of the higher to lower frequency QPO
appears to be about 3:2 and they have small amplitudes and coherences.
See Ref.~\refcite{michiel} for a review.

The frequencies of the two (`lower'  and `upper') kHz QPOs follow a tight relation for all neutron stars
from which they have been observed\cite{pbvdk}.
Most models for the generation of the kHz QPO pairs relate at least one of them to an orbital motion 
in the accretion disc. 
In the sonic-point spin-resonance model\cite{spsr}, the upper kHz QPO is 
generated by
a bright spot on the stellar surface generated by gas from a clump
orbiting at the sonic radius, 
whereas the lower kHz QPO is produced by illumination of gas at the spin-resonance radius.
In the relativistic precession model\cite{stella}, 
the upper and lower kHz QPO are identified respectively as the nodal and periastron precession of gas clumps at the inner edge of the disc.
Relativistic resonance models \cite{1235} suggest resonances between orbital frequencies in the disc, which would produce kHz QPO frequencies at a fixed ratio, similar to what is occasionally seen in black hole systems.

Circinus X-1 is an LMXB whose nature was long disputed. Although Type-I X-ray bursts were observed from the field of the object \cite{tennant87},
its X-ray, radio, and spectroscopic properties are more typical of a black hole.
Here we summarize the discovery of kHz QPO pairs reported in Ref.~\refcite{apj}. These show it is a neutron star.

\enlargethispage*{6pt}

\section{Observations}
We have analyzed  $\sim$2~Msec of archival data taken during 1996--2005 using the Rossi X-ray Timing Explorer
in the 3-60 keV energy range. We divided the observations into segments of 16, 48, 128, or
256 sec, Fourier-transformed them, and then averaged each one to get power spectra.
We made a precise estimate of the deadtime values of the detector by fitting the Zhang function\cite{zhang} to observations where Circinus X-1 does not seem to contribute, and used those to subtract the Poisson noise power from all observations.
After renormalizing the power spectra to rms squared we fitted a multi-Lorentzian function 
for the characteristic frequencies.
By plotting the frequencies of the fitted QPOs from each observation against each other as in 
Ref.~\refcite{steve}, we found two groups of correlated frequencies extending over several hundred Hz. 
Based on the high frequencies and the over an order of magnitude frequency variability we identified those as lower and upper kHz QPOs.
We found in total 80 observations with kHz QPOs. In 11 cases both appeared simultaneously.
All their characteristics can be found in Ref.~\refcite{apj}; in general they have relatively low coherences and amplitudes.

\begin{figure}[t]
\psfig{file=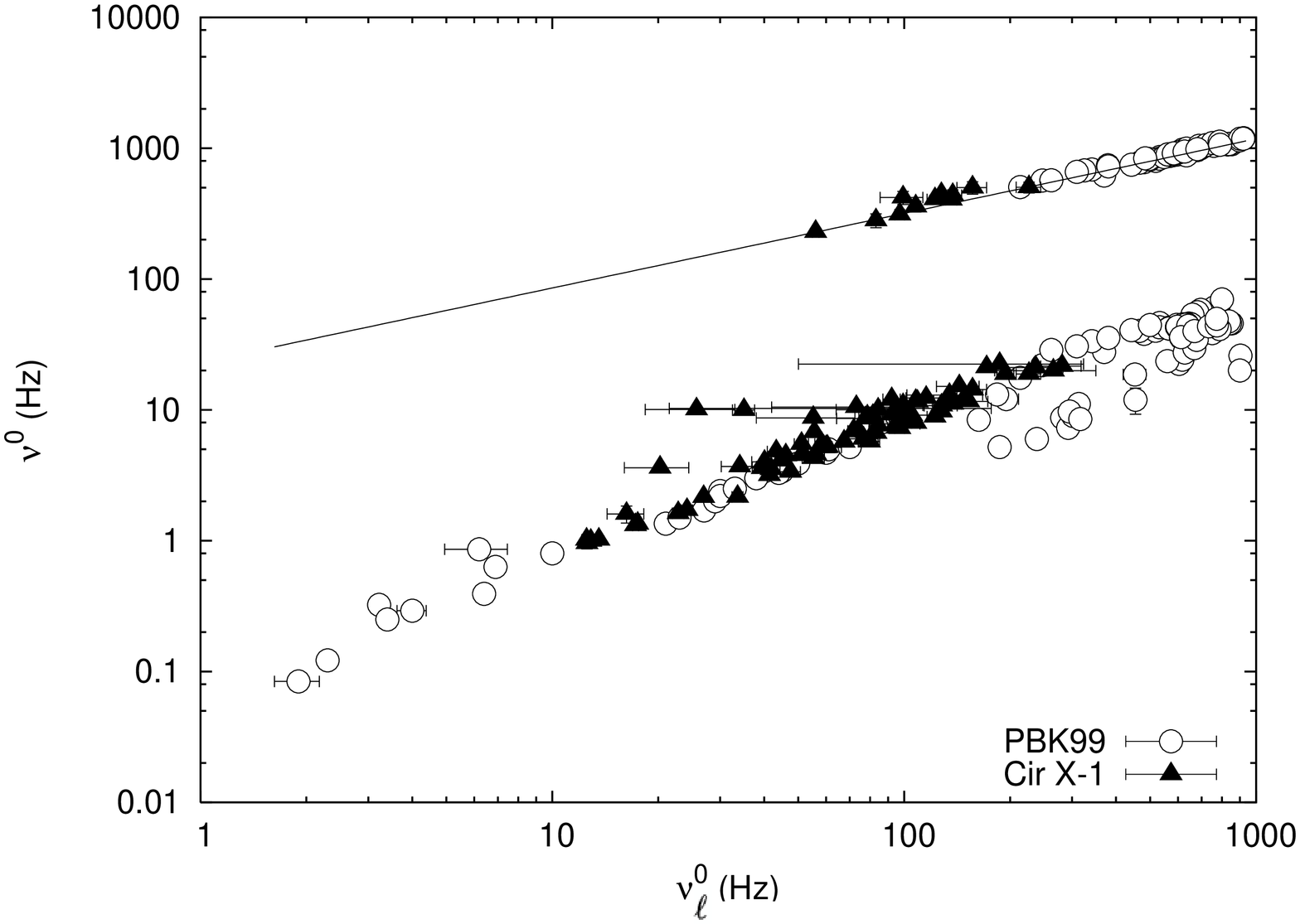,width=1.75in,angle=-90} 
\psfig{file=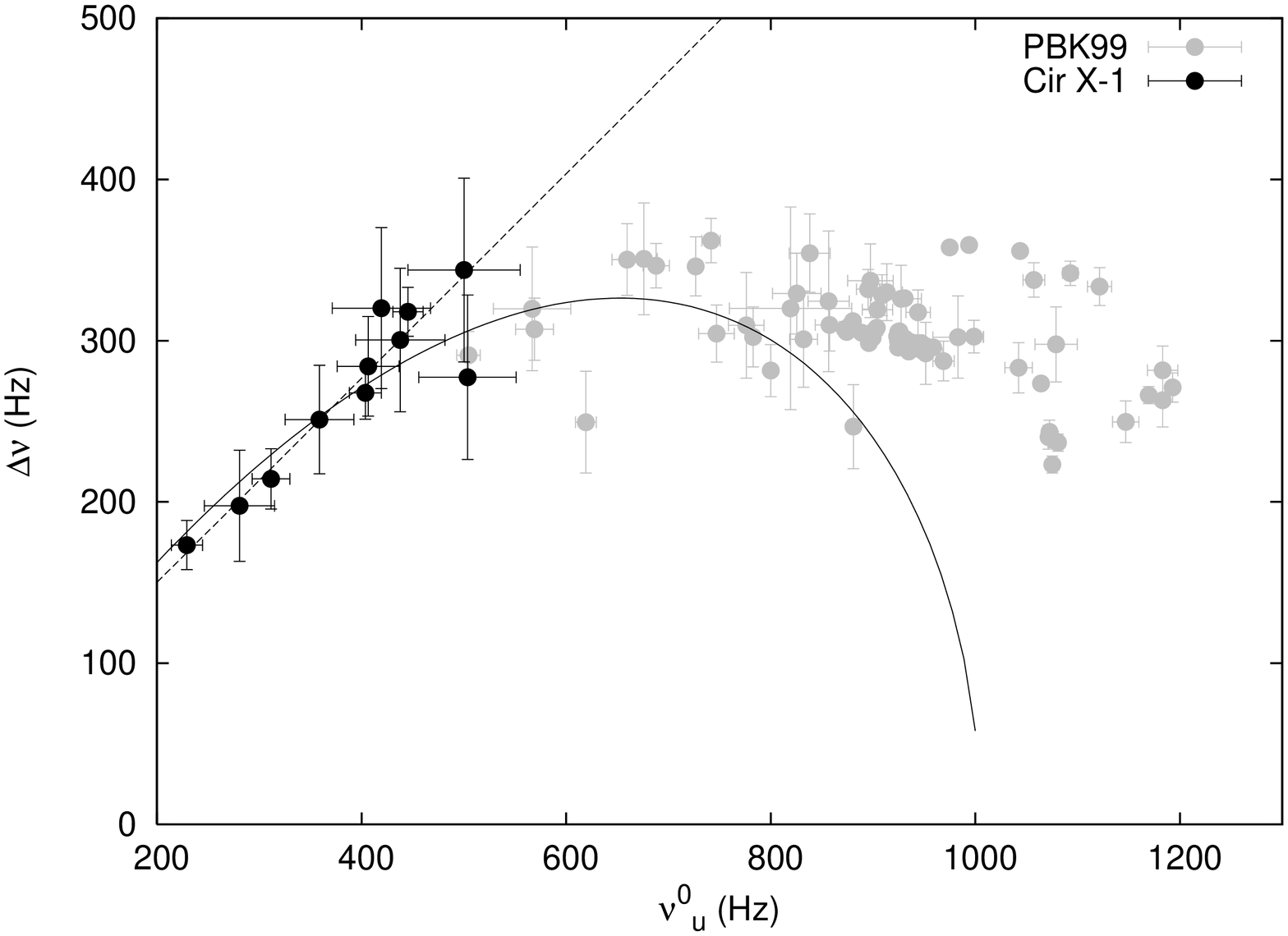,width=1.7in,angle=-90} 
\caption{Left: the frequencies of the upper kHz QPO (top) and of the low-frequency QPO (bottom) against that of the lower kHz QPO for our results from Circinus X-1 as also for other sources from Ref.~\refcite{pbvdk}. Right: The frequency separation of the kHz QPO pairs for the same sources as before, as well as the best fit for a straight line (with a slope of 0.6$\pm0.1$) and the relativistic precession model to the Circinus X-1 data. The slight misplacement of the latter in Ref.~\refcite{apj} is corrected here.}
\label{relations}
\end{figure}

In the left panel of Fig.~\ref{relations} we plot the two frequencies.
They follow the same relation that was found for other neutron stars
in Ref.~\refcite{pbvdk}, but 
extend this relation to frequencies lower by a factor 4.
It is of particular interest to see how the frequency separation behaves in this new frequency range for kHz QPOs.
We see in the right panel of Fig.~\ref{relations} that it increases over a factor 2 with increasing frequency of the lower kHz QPO. This is different from what is seen in other sources that have shown kHz QPO pairs at higher frequencies. Mathematically speaking, the relativistic precession model and to a lesser extent the relativistic resonance model, can explain the observed behavior for a neutron star mass of 2.2$\pm0.3 M_{\odot}$ or a 1:3 resonance respectively\footnote{Note that the line for the relativistic precession model is slightly different than in Ref.\refcite{apj} because of an error there in plotting the corresponding function.}.
The former is produced for slightly inclined, infinitesimally eccentric geodesic orbits around non-rotating compact objects, a framework that has been challenged \cite{markovic}.
Oscillations in the disk require an energy 
comparable to the binding energy of the star to produce relative amplitudes similar to those observed.
The sonic-point spin-resonance model can explain the generation of kHz QPOs, but is challenged by the strongly increasing frequency separation with kHz QPO frequency and the low frequencies observed in Cir X-1.

\section*{Acknowledgments}
This work was accomplished under the Marie-Curie Training Program HPMT-CT-2001-00245. S.B. acknowledges support by grants NSF AST 00-98399 and NASA NAG 5-12030 and the funds of the Fortner Endowed Chair at the University of Illinois.

\vfill

\end{document}